\documentclass[lettersize,journal]{IEEEtran}
\usepackage{amsmath,amsfonts}
\usepackage{amssymb}
\usepackage{algorithmic}
\usepackage{algorithm}
\usepackage{array}
\usepackage[caption=false,font=normalsize,labelfont=sf,textfont=sf]{subfig}
\usepackage{textcomp}
\usepackage{stfloats}
\usepackage{url}
\usepackage{verbatim}
\usepackage{graphicx}
\usepackage{cite}
\usepackage{orcidlink}
 \usepackage{balance}
\usepackage{multirow}
\usepackage{multicol}
\newcolumntype{d}[1]{D..{#1}}
\usepackage{multirow}
\usepackage{amssymb}
\usepackage{booktabs}
\usepackage{booktabs}
\usepackage{tcolorbox}
\usepackage{flushend}
\usepackage{makecell}
\usepackage{colortbl}
\usepackage{color}
\usepackage{tablefootnote}
\usepackage{threeparttable} % to use table notes
\usepackage{tikz}
\usepackage{graphicx}
\usepackage{subcaption}
\usetikzlibrary{positioning, arrows.meta}

\begin{document}

\title{Libri2Vox Dataset: Target Speaker Extraction with Diverse Speaker Conditions and Synthetic Data}

\author{Yun Liu , \IEEEmembership{Student Member, IEEE}, Xuechen Liu , \IEEEmembership{Member, IEEE}, Xiaoxiao Miao , \IEEEmembership{Member, IEEE}, and Junichi Yamagishi, \IEEEmembership{Senior Member, IEEE}
\thanks{This study is partially supported by MEXT KAKENHI Grants (24K21324) and JST, the establishment of university fellowships towards the creation of science technology innovation (JPMJFS2136).}
\thanks{Yun Liu, Xuechen Liu and Junichi Yamagishi are with National Institute of Informatics, Tokyo 101-8430, Japan (e-mail: \{yunliu, xuecliu, jyamagis\}@nii.ac.jp). Xiaoxiao Miao is with Singapore Institute of Technology, 10 Dover Dr, Singapore 138683 (e-mail: xiaoxiao.miao@singaporetech.edu.sg).}
}

% The paper headers
\markboth{Journal of \LaTeX\ Class Files,~Vol.~XX, No.~X, XXX~2022}%
{Shell \MakeLowercase{\textit{et al.}}: The PartialSpoof Database and Countermeasures for the Detection of Short Generated Audio Segments Embedded in a Speech Utterance}

\maketitle

\begin{abstract}

Target speaker extraction (TSE) is essential in speech processing applications, particularly in scenarios with complex acoustic environments. Current TSE systems face challenges in limited data diversity and a lack of robustness in real-world conditions, primarily because they are trained on artificially mixed datasets with limited speaker variability and unrealistic noise profiles. To address these challenges, we propose Libri2Vox, a new dataset that combines clean target speech from the LibriTTS dataset with interference speech from the noisy VoxCeleb2 dataset, providing a large and diverse set of speakers under realistic noisy conditions. We also augment Libri2Vox with synthetic speakers generated using state-of-the-art speech generative models to enhance speaker diversity. Additionally, to further improve the effectiveness of incorporating synthetic data, curriculum learning is implemented to progressively train TSE models with increasing levels of difficulty. 
Extensive experiments across multiple TSE architectures reveal varying degrees of improvement, with SpeakerBeam demonstrating the most substantial gains: a 1.39 dB improvement in signal-to-distortion ratio (SDR) on the Libri2Talker test set compared to baseline training. Building upon these results, we further enhanced performance through our speaker similarity-based curriculum learning approach with the Conformer architecture, achieving an additional 0.78 dB improvement over conventional random sampling methods in which data samples are randomly selected from the entire dataset.
These results demonstrate the complementary benefits of diverse real-world data, synthetic speaker augmentation, and structured training strategies in building robust TSE systems.

\end{abstract}


\section{Introduction}
Target speaker extraction (TSE) \cite{zmolikova2023neural} is a key task in speech processing, focusing on isolating the voice of a desired speaker from complex acoustic environments. This capability is valuable in applications such as voice-controlled systems, teleconferencing, and hearing aids, where extracting clear speech signals directly impacts system performance and user experience. Despite notable advances, TSE still faces multiple challenges, particularly related to limited data diversity and a lack of robustness under real-world conditions~\cite{zmolikova2023neural}.

A significant issue with TSE is the mismatch between training and deployment environments. Current models are typically trained on artificially mixed speech, which is controlled but fails to capture the complexity of real-world conditions~\cite{wang2024speech}. The diverse nature of noise, including spatial configurations, reverberations, and dynamic changes, leads to significant performance degradation during deployment. The limitations of current TSE datasets (highlighted in Table~\ref{tab:3dataset}) are evident in both  restricted speaker diversity and the controlled, synthetic nature of mixtures. For instance, datasets such as WSJ0-2mix-extr~\cite{xu2020spex,ge2020spex+} and Libri2talker~\cite{xu2021target} are limited regarding the number of speakers, with 101 and 1172, respectively. Moreover, these datasets possess limited variability in terms of acoustic and speaker conditions. These limitations can lead to models that generalize poorly to unseen speakers and complicated  real-world  acoustic environments. Addressing these challenges and improving the robustness of TSE systems thus requires incorporating more variations in terms of noise conditions and speakers.

To address these limitations and enhance TSE system generalization, we propose a novel data integration approach focusing on two critical aspects: speaker diversity and acoustic variability.  We leverage the VoxCeleb2 dataset~\cite{chung2018voxceleb2}, which encompasses more than 6,000 speakers recorded in diverse acoustic environments, providing a rich source of real-world variations. However, directly using the noisy recordings from VoxCeleb2 as target speech contradicts the objective with TSE, which is to extract clean speech, thereby degrading the effectiveness of the training data. To resolve this constraint while maintaining data diversity, we implement a strategic combination: employing the high-fidelity LibriTTS dataset~\cite{zen2019libritts}, derived from LibriSpeech~\cite{panayotov2015librispeech}, as our target speaker source while utilizing VoxCeleb2 for interference speaker source.

Another solution to address such challenges is the acquisition of synthetic data, which has demonstrated remarkable efficacy across various alternative tasks~\cite{lu2023machine,nikolenko2021synthetic}, such as computer vision~\cite{Mayer2018What,Esteva2021Deep} and natural language processing~\cite{Fadaee2017Data,Feng2021A,wang2024survey}. It has been a promising solution on data scarcity and model robustness. 
Synthetic data have also proven highly effective in various speech-related tasks. In automatic speech recognition (ASR), for example, the use of synthetic speech data generated using text-to-speech (TTS) models has demonstrated considerable improvements~\cite{hu2021synt++,fazel2021synthasr}. Similarly, research on TTS systems has shown that training robust models on synthetic data produced using less stable systems can enhance transfer stability, delivering high-quality transfers while retaining speaker characteristics~\cite{shen2019effective,chen2020tts}. For speaker-related tasks, synthetic data also enables TTS systems to synthesize speech from new, unseen speakers by sampling from a learned latent distribution~\cite{li2018speaker,zhang2018overcoming}.
Building upon these advances, we extend synthetic data approaches to TSE. Our goal is to enable TSE models to reliably extract clean speech in diverse, real-world environments. Incorporating synthetic data helps address challenges of data scarcity, especially in handling new speakers and complex acoustic conditions.

Our two previous studies laid the foundation for this study. In our initial work~\cite{liu2024target},  we implemented a curriculum learning (CL) approach~\cite{soviany2022curriculum} that progressively trains the model with increasingly complex scenarios, demonstrating significant performance improvements in challenging speaker extraction tasks. Our subsequent research~\cite{liuslt}, we introduced the use of synthetic speakers to the learning scheme. The synthetic speakers were generated through voice conversion and speaker anonymization~\cite{Lv2023SALTDS}, which led to substantial improvements in TSE performance. 
Voice conversion transforms a source speaker into a target speaker, producing speech with a different identity. Speaker anonymization, similarly, takes speech in but removes the speaker's identity, creating a new, anonymized speaker that does not exist in reality.
In this study, we extended previous efforts by using VoxCeleb2, a larger dataset with more speakers than current TSE dataset, which enables the generation model to generate a greater number and variety of synthetic speakers.

The main contributions of this study are threefold:
\begin{itemize}
    \item \textbf{New dataset called Libri2Vox}:
    Libri2Vox includes a large and diverse set of speakers with realistic noisy interference. The target speakers are sourced from the cleaner LibriTTS corpus, while interference speakers are derived from VoxCeleb2. This combination allows for better representation of real-world acoustic conditions while maintaining clean target speech signals.

    \item \textbf{Libri2Vox variants with synthetic speakers}:
    We further expand the diversity of Libri2Vox by incorporating synthetic speakers generated using  speech generative models. These synthetic speakers introduce additional variability, which is expected to enhance the robustness and generalization capability of TSE models. Our previous study demonstrated that increasing the diversity of interference speakers through synthetic generation can significantly enhance TSE performance, particularly when combined with CL~\cite{liuslt}. In this study, we built upon these findings by using the more diverse VoxCeleb2 dataset, which provides a richer pool of speakers for generating synthetic data.

    \item \textbf{Investigating effectiveness of synthetic speakers in curriculum learning}:
    We also explored the impact of synthetic speakers on TSE model performance by incorporating CL. Our experiments revealed that synthetic speakers with CL can be used to significantly boost the performance of models trained on Libri2Vox,  highlighting the value of synthetic data in progressively enhancing speaker extraction capabilities. 
\end{itemize}

\begin{figure}[t]
    \centering
    \includegraphics[width=0.6\columnwidth]{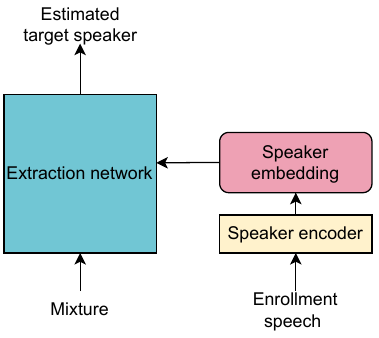}
    % \vspace{-1cm}
    \caption{Basic conceptual framework of TSE.}
    \label{fig:tse}
\end{figure}
% \vspace{-1cm}

\section{Related work}
This section defines TSE, discusses deep neural network (DNN)-based TSE models, provides an overview of popular datasets while identifying their shortcomings, and presents effective training strategies.
\subsection{Target speaker extraction definition}
\label{sec:PS}

% The basic framework of the task of TSE is illustrated in Fig. ~\ref{fig:tse}. Mathematically, it can be expressed as below:
% \[
% \hat{\mathbf{w}}^{(t)} = \text{TSE}(m(t), e_t = E(\mathbf{w}^{(r_t)}); \theta),
% \]
% where given a reference waveform $\mathbf{w}^{(r_t)} = (w_1, \cdots, w_{r_t})$ containing speech signals of the target speaker $t$, this waveform is used to extract a speaker embedding $e_t = E(\mathbf{w}^{(r_t)})$ via a neural speaker encoder $E$. TSE aims to output the estimated clean speech signals $\hat{\mathbf{w}}^{(t)} = (\hat{w}_1, \cdots, \hat{w}_t)$ of the target speaker from a given mixture $m = (m_1, \cdots, m_t)$, where $m=s+s'$ contains the target speaker’s clean speech $s$ and interference speakers' signals $s'$. The notation $\theta$ represents the model parameters of the extraction framework.

The basic framework of the task of TSE is illustrated in Fig. ~\ref{fig:tse}. Mathematically, it can be expressed as below:
\[
\hat{\mathbf{w}}^{(t)} = \text{TSE}(m, e_t = E(\mathbf{w}^{(r_t)}); \theta),
\]
where given a enrollment waveform $\mathbf{w}^{(r_t)}$ containing speech signals of the target speaker $t$, this waveform is used to extract a speaker embedding $e_t = E(\mathbf{w}^{(r_t)})$ via a neural speaker encoder $E$. TSE aims to output the estimated clean speech signals $\hat{\mathbf{w}}^{(t)} $ of the target speaker from a given mixture $m$, where $m=s+s'$ contains the target speaker's clean speech $s$ and interference speakers' speech $s'$. The notation $\theta$ represents the model parameters of the extraction framework.

\subsection{Deep neural networks for target speaker extraction}
DNNs have enabled  significant advances on developing single-channel TSE systems. Early approaches primarily utilized a reference signal from the target speaker to guide the extraction process, differing in network architectures and the manner in which speaker embeddings were used. The following DNNs are the most common for TSE:

SpeakerBeam \cite{zmolikova2019speakerbeam} was the first model specifically designed for TSE. Unlike other speech separation models that attempt to determine the number of speakers in a mixture, SpeakerBeam focuses exclusively on extracting the target speaker. By leveraging speaker information through embeddings, it effectively isolates the desired speaker's voice, overcoming common issues such as label permutation and speaker tracing. Notably, SpeakerBeam's speaker embedding is extracted using a simple speaker encoder  trained jointly with the TSE network.

VoiceFilter \cite{wang2019voicefilter} integrates convolutional layers, long-short term memory (LSTM), and fully connected layers. Unlike SpeakerBeam, speaker embeddings used by VoiceFilter are extracted from a pre-trained speaker encoder ~\cite{variani2014deep}, providing fixed guidance.

% SpeakerFilter~\cite{he2020speakerfilter} introduces improvements that effectively utilize reference speech through a multi-level feature extraction process at different resolutions and seamlessly integrating these features into a UNet \cite{unet}.

The Conformer~\cite{gulati2020conformer} architecture combines convolutional layers and self-attention mechanisms to simultaneously capture local and global dependencies in speech input. The Conformer-based TSE model~\cite{liu2024target} processes the time-frequency domain short-term Fourier transform (STFT) spectrum of the input mixture. By using a Conformer block—a hybrid of multi-head self-attention and full convolution—Conformer generates the real and imaginary parts of the target speech signal's STFT. Conformer blocks also integrate STFT features with speaker embeddings extracted from reference utterances using a pre-trained speaker encoder \cite{desplanques2020ecapa}. Unlike SpeakerBeam and VoiceFilter, which enhance only the magnitude and use the noisy phase for reconstruction,  Conformer predicts the complex spectrum mask~\cite{williamson2016complex}.

We use all of these models to prove the effectiveness of Libri2Vox and our training strategy. 

\begin{table}[t]
\centering
% \scriptsize
\caption{Comparison of WSJ0-2mix-extr, Libri2mix, and Libri2talker Datasets.}
\begin{tabular}{|l|c|c|c|}
\hline
\textbf{Dataset}         & \textbf{\# Speakers} & \textbf{\# Utterances}   & \textbf{Duration(h)} \\ \hline
\multirow{3}{*}{\textbf{WSJ0-2mix-extr}}  & 101 (train) & 20,000 (train) & 30 \\  
 & 101 (val) & 5,000 (val) & 8 \\  
 & 18 (test) & 3,000 (test) & 5 \\ \hline
\multirow{3}{*}{\textbf{Libri2mix}}       & 921 (train-360) & 50,800 (train-360) & 212 \\  
 & 251 (train-100) & 13,900 (train-100) & 58 \\  
 & 40 (val) & 3,000 val & 11 \\ 
  & 40 (test) & 3,000 test & 11 \\ \hline
\multirow{3}{*}{\textbf{Libri2talker}}    & 1,172 (train/val) & 127,056 (train) & 460 \\  
 & 1,172 (val) & 2,344 (val) & 8 \\  
 & 40 (test) & 6,000 (test) & 22 \\ \hline
\end{tabular}
\label{tab:3dataset}
\end{table}

\subsection{Datasets for target speaker extraction}
\label{sec:Datasets}
Table~\ref{tab:3dataset} presents three major datasets: WSJ0-2mix-extr, Libri2mix, and Libri2talker. These datasets are described in this section.

WSJ0-2mix-extr \cite{xu2020spex} is derived from the WSJ0~\cite{garofolo1993csr} corpus, which consists of clean, read speech from the Wall Street Journal. For each 2-talker mixture audio in WSJ0-2mix-extr, two randomly selected utterances from different speakers in WSJ0 are mixed. This dataset is composed of training, development, and evaluation sets, with the training set including 20,000 mixtures generated from 101 speakers (50 male and 51 female),  development set containing 5,000 mixtures, and  evaluation set containing 3,000 mixtures involving 18 different speakers not seen during training. The signal-to-noise ratio (SNR) of these mixtures was chosen between 0  and 5 dB. Most of the 2-talker speech samples are heavily overlapped, creating a challenging environment for TSE models.

Libri2mix~\cite{cosentino2020librimix} is derived from LibriSpeech \cite{panayotov2015librispeech}. It creates clean 2-talker mixtures by randomly selecting and mixing two utterances from its predecessor. It consists of several subsets, including train-360 with 50,800 utterances (212 hours, 921 speakers) and train-100 with 13,900 utterances (58 hours, 251 speakers), as well as dev and test sets, each containing 3,000 utterances (11 hours, 40 speakers). The dataset follows a minimum duration protocol, trimming the longer utterance in a pair to match the shorter one, resulting in a 100\% overlap rate. This setup is designed to provide challenging conditions for speech separation models.

Libri2talker, an extended version of Libri2mix, reuses 2-talker mixtures by swapping the roles of the target and interference speakers, effectively doubling the available data. It includes a training set with 127,056 examples from 1,172 speakers,  validation set with 2,344 examples, and  evaluation set with 2,260 enrollment utterances from the LibriSpeech test-clean set and 6,000 test utterances from the Libri2mix test set. Like Libri2mix, Libri2talker applies the minimum duration protocol and retains a 100\% overlap rate, making it suitable for both speaker verification and TSE tasks.

These datasets are composed of artificially mixed speech, where both speakers in each mixture were recorded under studio conditions, resulting in clean speech. This setup leads to a significant mismatch with real-world data, which typically contains more variability and noise. Since our ultimate goal is to handle real-world scenarios, using real-world data as interference sources is a more suitable approach to bridge this gap and enhance model robustness.

\subsection{Training strategies for separation-related task}

There are predominately two categories of state-of-the-art training strategies for TSE: data simulation and optimization strategies. Data simulation is crucial for augmenting mixture training data, particularly when real-world labeled datasets are limited. Common techniques include:

\begin{itemize}
    \item \textbf{Data Augmentation~\cite{alex2023data}}: This involves generating new training data by modifying existing datasets using techniques such as adding Gaussian noise, pitch shifting, or time stretching. Data augmentation helps train more robust models that generalize better to different environments.
    \item \textbf{Dynamic Mixing}: This strategy dynamically generates new mixtures of target and interference speakers during training. By continuously varying the conditions in which the target speaker is extracted, dynamic mixing improves the model's generalization to different noise and interference scenarios.
\end{itemize}
Optimization strategies have also demonstrated significant promise in improving TSE performance. Common strategies include:
\begin{itemize}
    \item \textbf{GAN-based Method}: 
To address the training/inference mismatch in deep noise suppression models, real data can be implemented using either generative models or reference-free loss without clean speech access \cite{realnoisy2024xu}. With this strategy, an end-to-end non-intrusive DNN, called PESQ-DNN, is used to estimate the perceptual evaluation of speech quality (PESQ)~\cite{rix2001perceptual} scores, providing a reference-free perceptual loss during training. An alternating training protocol is applied, where the DNS model is updated on real data, followed by PESQ-DNN updates on synthetic data. This strategy significantly improves the performance compared with models trained solely on simulated data.

    \item \textbf{Curriculum Learning}: 
     CL is a strategy with which training data is introduced progressively, starting with easier examples and moving toward more complex ones. CL has been implemented in TSE by sorting training samples on the basis of predefined difficulty measures such as gender, speaker similarity, signal-to-distortion ratio (SDR), and SNR. Initially, easier samples in which the target and interfering speakers are more distinct are used, and progressively harder cases are introduced as training advances. CL has been shown to improve model convergence and performance~\cite{liuslt,liu2024target} and were used in our experiments to optimize TSE training.

\end{itemize}

\begin{figure}[t]
    \centering
    \includegraphics[width=0.8\columnwidth]{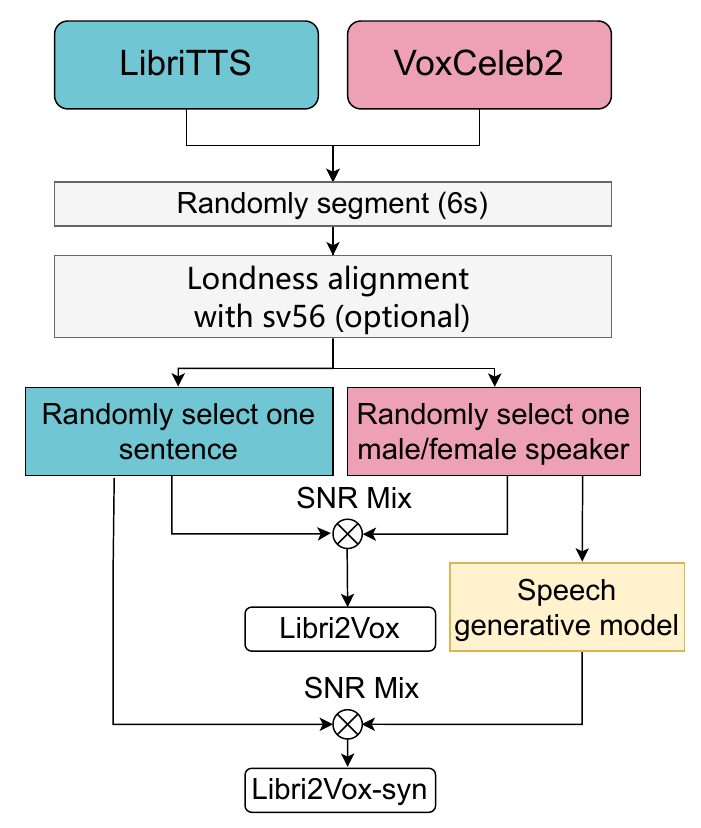}
    % \vspace{-1cm}
    \caption{Data generation framework of Libri2Vox and its synthetic version.}
    \label{fig:datagen}
\end{figure}

\section{Libri2Vox Dataset}
\label{sec:database}

TSE systems face two critical challenges: limited speaker diversity and artificial acoustic environments. Current TSE datasets, as shown in Table \ref{tab:3dataset}, typically contain only hundreds to a few thousand speakers, significantly constraining model generalization capabilities. This limitation becomes particularly apparent when systems encounter speakers or acoustic conditions outside the training distribution.

Speaker diversity plays a fundamental role in TSE performance. A rich speaker set enables models to learn robust representations across various speech characteristics, including accent variations, speaking styles, and vocal qualities. Our analysis suggests that expanding the speaker sets directly correlates with improved model generalization and performance metrics.

Apart from speaker diversity, acoustic condition is another intriguing factor. Most existed datasets for standard TSE training and evaluation are generated by artificially adding target and interference speech with noise, which does not fully capture how noise and speech mixtures occur in real environments. The complexity of how background noise interacts with speech, including varying distances between the speakers and noise sources, or overlapped voices during conversations, makes it difficult to simulate the true nature of how speech mixtures are created in real-time cases. This is where current datasets often fall short. Meanwhile, datasets that have been widely used  for other tasks (such as VoxCeleb2 used in this study) often contain real-world recordings, some of which include background noise from various environments, making them a closer reflection of the conditions under which TSE models are expected to operate.

To address these limitations, we leverage VoxCeleb2 recordings as interference sources. These recordings inherently contain diverse, real-world acoustic conditions, providing authentic training scenarios for TSE models. This approach marks a significant departure from traditional artificial mixing approaches while maintaining the controlled conditions necessary for systematic model development.

\subsection{Data construction}
\label{sec:database-procedures}
In constructing the \textit{Libri2Vox} database, as shown in Fig. \ref{fig:datagen}, we used several following key steps.
%to ensure high-quality audio data mixing and the generation of synthetic interference speakers.

\subsubsection{Pre-processing and mixing}
Each audio segment is randomly split into 6-second segments. Segments shorter than six seconds are zero-padded at the end. The dataset is first processed in the following steps:

\begin{itemize}
    \item All audio data is pre-processed using \textbf{sv56}\footnote{https://github.com/foss-for-synopsys-dwc-arc-processors/G722/tree/master/sv56} with a scale factor of -26 dB. 
    \item In \textit{LibriTTS}, target speech shorter than 2s is deleted.
    Then, speakers with fewer than 3 utterances are removed, resulting in the deletion of 31 speakers out of 1,151, leaving 1,120 speakers in total.
    \item For reference speech longer than 10s, segments are randomly concatenated, ensuring no segment exceeds 15s.
\end{itemize}

For the dataset composition, we use VoxCeleb2 as the interference speaker dataset and LibriTTS as the target speaker dataset. 
VoxCeleb2 is a large-scale speaker recognition dataset that was collected "in the wild", meaning that the speech segments are naturally corrupted by real-world noise such as laughter, cross-talk, channel effects, music, and other background sounds. This adds an element of realism, making the dataset particularly valuable for training models that need to handle noisy environments. VoxCeleb2 is also multilingual, featuring speech from speakers of 145 different nationalities and covering a wide range of accents, ages, and languages. 
One potential problem with VoxCeleb2 for this study is that the speech segments are predominantly recorded in noisy conditions. However, on the other hand, the inherent noise in VoxCeleb2 makes it an ideal source for interference speakers. By using VoxCeleb2 as the interference speaker source, we can take advantage of the real-world noise captured during the original recordings, rather than relying on artificially added noise in post-processing. This enables the dataset to more accurately simulate real acoustic conditions in TSE.

VoxCeleb2 includes both a development (dev) set and  evaluation set. The dev set contains 5,994 speakers, with a total of 1,092,009 utterances, while the evaluation set includes 118 speakers and 36,237 utterances. The entire dataset contains over 1.1 million utterances, making it a rich resource for training models on diverse speaker characteristics. The average duration of the utterances in VoxCeleb2 is approximately 7.8s. We randomly select 94 speakers from the dev set of VoxCeleb2 to form the validation set and leave 5,900 speakers for training.

The other data source acquired for this study is LibriTTS \cite{zen2019libritts}. It is a refined version of LibriSpeech \cite{panayotov2015librispeech}, where a small portion of the noisy utterances from the latter was removed to ensure cleaner speech for tasks such as TTS.

For each target speech from LibriTTS, we randomly select one male and one female interference speaker from VoxCeleb2 in an alternating manner. This means that for every target speech, we first choose a male interference speaker then a female interference speaker, ensuring a balanced gender distribution across the dataset. The mixture is then created by adding the target speech to the interference speech with an additional scaling factor $\alpha$, which is sampled uniformly from the SNR range of [-5, 5] dB.  The same mixing procedure applies for the validation and test sets, with the average SNR across the train, validation, and test sets being approximately -0.11 dB.

\subsection{Statistics of   Libri2Vox}

\textit{Libri2Vox} was pre-split after collecting and processing, as shown in Table \ref{tab:libri2vox_stats}
For the training set,  \textit{LibriTTS} provides 1,151 speakers, with 8.97 hours of data. Each \textit{LibriTTS} utterance generates one corresponding data point (mixture, reference, target). VoxCeleb2 contributes  5,900 interference speakers for this partition, making the training set contains 149,691 triplets, equivalent to 250 hours of mixture data.
The validation set consists of 40 \textit{LibriTTS} speakers (approximately 8.97 hours) and 94 VoxCeleb2 speakers, totaling 118 speakers and 7,200 utterances.
The test set includes 39 \textit{LibriTTS} speakers (approximately 8.56 hours) and 118 VoxCeleb2 speakers, with a total of 157 speakers and 6,000 utterances.

\begin{table}[t]
\centering
\caption{Libri2Vox statistics.  "Total" column shows the total number of utterances combined with corresponding  LibriTTS and VoxCeleb2 sets.}
\begin{tabular}{|l|c|c|c|c|}
\hline
\multirow{2}{*}{\textbf{Set}} & \multicolumn{2}{c|}{\textbf{\#. Speakers}} & \textbf{\#. Utterances} & \multirow{2}{*}{\textbf{Duration (h)}} \\ \cline{2-4}
& \textbf{LibriTTS} & \textbf{VoxCeleb2} & \textbf{Total} &  \\ \hline
\textbf{Training} & 1,151 & 5,900 & 149,691 & 250  \\ \hline
\textbf{Validation} & 40 & 94 & 7,200 & 8.97  \\ \hline
\textbf{Testing} & 39 & 118 & 6,000 & 8.56  \\ \hline
\end{tabular}
\label{tab:libri2vox_stats}
\end{table}

\section{Synthetic Libri2Vox Dataset }
\label{sec:overview_CM}

\subsection{Why synthetic speakers?}
\label{sec:overview-task}
Enhancing the diversity of training data can significantly improve the performance of TSE models. Conventionally, this has been achieved by applying data augmentation to real data~\cite{alex2023data}. However, the diversity provided by data augmentation is limited in terms of the range of data distribution it can cover. Another approach to generating large amounts of data with diverse speaker characteristics is to use speech generative models~\cite{ju2024naturalspeech3,wang2023valle,ren2021fastspeech2}. These recent models enable speech generation with a high level of naturalness and speaker similarity. This development brings up a key question: can these generative models be used to generate specialized training data for TSE?

There are several strategies to generate training data with generative models, one of which involves producing diverse synthetic interference speakers from the existing interference ones,  ensuring the defined difference~\cite{Miaosynvox2,Lv2023SALTDS}. This is the strategy we investigated.

\subsection{Two types of synthetic interference speaker generation methods}
We introduce two different generation methods used to generate synthetic interference speakers that are distinct from the real speakers selected from VoxCeleb dataset, i.e. SynVox2~\cite{Miaosynvox2}  and speaker anonymization through latent transformation (SALT)~\cite{Lv2023SALTDS}. These two methods are used to increase the diversity of interference speakers used in TSE tasks. Below is a description of \textbf{SALT} .

\subsubsection{SALT}
SALT generates synthetic interference speakers by manipulating the latent space of pre-trained speaker representations. In the context of speaker extraction, let $s_i$ represent the given interference speaker's audio. The steps to generate a synthetic speaker with  SALT  are as follows:

\begin{itemize}
    \item \textbf{WavLM Representation Extraction}: We first extract the WavLM \cite{chen2021wavlm} representation, $\text{WavLM}(s_i)$, of the input interference speaker speech $s_i$. This representation captures both the content and speaker-related features, as evident in that previous study~\cite{chen2021wavlm}.
    Then, reference speaker($s_r$) representations, $\text{WavLM}(s_r^j),j \in [1, N]$, are also extracted from a pool of $N$ speakers by WavLM.
    
\item \textbf{$\textbf{k}$-nearest neighbor (k-NN) Search}: Given $\text{WavLM}(s_i)$ as the query, a $k$-nearest neighbor (k-NN)~\cite{knn} search is conducted for each frame of the query representation. At each time step, the $k$ most similar frames are selected from the reference speaker representations of all $N$ speakers. This process yields a set of closest representations $\mathbf{D}_j$ for each of the $N$ selected reference speakers.

\item \textbf{Weighted Summation}: After selecting $k$-NN representations, random weights $w_j$ are assigned to each of the reference speaker representation sets, $\mathbf{D}_j$. These weights are sampled from a normal distribution and normalized to sum to one. The weighted sum of the selected representations is then combined with the original interference speaker representation via linear interpolation, as follows:
\begin{align*}
    (\mathbf{D}_1, \cdots, \mathbf{D}_N) & = \text{kNN} (\text{WavLM}(s_i), \text{WavLM}(s_r^j)) \\ 
    \mathbf{O} & = (1 - p) \sum_{j=1}^{N} w_{j} \mathbf{D}_j + p \cdot \text{WavLM}(s_i)
\end{align*}

    where $p$ is a parameter that controls the balance between the original and the synthetic representations.

    \item \textbf{Vocoder-based Reconstruction}: Finally, the interpolated representation $\mathbf{O}$ is passed through the HiFi-GAN vocoder~\cite{kong2020hifi} to generate a waveform of the synthetic interference speaker.
\end{itemize}
For this study, the number of nearest neighbors considered for interpolation was $k = 4$, and the interpolation weight between the original speaker representation and  synthetic speaker representation was $p = 0.5$.
%
%For the reference pool, 
We used the \texttt{WavLM-Base}\footnote{\url{https://huggingface.co/microsoft/wavlm-base}} model trained on LibriSpeech, to extract latent space representations from the third layer. For the interpolation method, 50 speakers are randomly chosen from the LibriSpeech train-clean-100 dataset. For each of these speakers, 50 audio samples are selected at random to extract their features. The number of random reference speakers, 
$N$, is fixed at 4. The setting are the same as the original paper~\cite{Lv2023SALTDS}.

By using SALT, we can generate a large variety of synthetic interference speakers that are different from real speakers, providing a more challenging dataset for TSE model training. This method enables for a balance between speaker similarity and diversity through parameters $k$ and $p$, which enables better control over the generated synthetic speakers.

\subsubsection{SynVox2}
SynVox2 was designed for speaker anonymization using an orthogonal Householder neural network (OHNN) \cite{10244064,Miaosynvox2}. The framework operates through the following three essential components:

\begin{itemize}
\item \textbf{Disentanglement:} Speech characteristics are derived using specialized encoders. The Yet Another Algorithm for Pitch Tracking (YAAPT) \cite{Kasi2002} extracts the fundamental frequency (F0), while the ECAPA-TDNN \cite{desplanques2020ecapa} speaker encoder, trained on VoxCeleb2, generates 192-dimensional speaker identity embeddings. Additionally, a Hidden Unit BERT (HuBERT)-based soft content encoder, fine-tuned on LibriTTS-train-clean-100 from a pre-trained HuBERT model \cite{Hsu2021}, captures linguistic content information.

\item \textbf{Anonymization:} The system employs an OHNN-based anonymizer \cite{10244064} that transforms original speaker embeddings into anonymized representations through multiple orthogonal Householder transformation layers. The network utilizes randomly initialized weights and is optimized using classification and distance-based loss functions to ensure the anonymized speaker identities are distinct from both the original and other anonymized speakers.

\item \textbf{Generation:} The synthesis stage integrates the extracted content features, F0 information, and anonymized speaker embeddings into a HiFi-GAN model trained on LibriTTS-train-clean-100. This integration produces high-quality anonymized speech waveforms that preserve natural speech characteristics while ensuring distinct speaker identities.
\end{itemize}

% SynVox2 is generated for speaker anonymization by utilizing an orthogonal Householder neural network (OHNN) \cite{10244064,Miaosynvox2}. The process involves  the following three key steps:
% \begin{itemize}
% \item \textbf{Disentanglement:} The  Yet Another Algorithm for Pitch Tracking(YAAPT)  algorithm ~\cite{Kasi2002} is applied to extract the fundamental frequency (F0). The ECAPA-TDNN~\cite{desplanques2020ecapa} speaker encoder, trained on VoxCeleb2 , provides 192-dimensional speaker identity representations. The  Hidden Unit BERT (HuBERT) based soft content encoder is fine-tuned on the LibriTTS-train-clean-100  dataset, starting from a pre-trained HuBERT-based model~\cite{Hsu2021}, to capture the linguistic content of speech.

% \item  \textbf{Anonymization:} The OHNN-based anonymizer \cite{10244064} rotates the original speaker embeddings into anonymized embeddings by applying several orthogonal householder transformation layers. The OHNN weights are randomly initialized and trained using classification and distance-based losses to ensure that the anonymized speakers do not overlap with the original or other anonymized speakers. Therefore, the anonymized embeddings generated using the OHNN-based anonymizer form distinct pseudo-speaker identities.
% \item \textbf{Generation:} Finally, the content features, F0, and anonymized speaker embeddings are combined and passed into a HiFi-GAN model  to generate the audio waveforms. The HiFi-GAN model is trained using the LibriTTS-train-clean-100 database .

% \end{itemize}

\subsection{Synthetic data and statistics}

Since the synthetic data generated for Libri2Vox-syn only slightly alters the duration of the original VoxCeleb2 recordings, and we randomly select 6-second audio segments for all speakers, the statistics for the Libri2Vox-syn dataset remain identical to those of the original Libri2Vox dataset. This means that the number of speakers,  total number of utterances, and the total duration of the dataset are consistent across both the real and synthetic versions.

\section{Constructing TSE Models on the Libri2vox Dataset}
\label{sec:proposed-cm}

\subsection{Architecture of different target speaker extraction models}
\label{sec:4-tse-NETWORK}
This section describes the four TSE neural networks we used for the experiments: Conformer \cite{gulati2020conformer}, VoiceFilter \cite{wang2019voicefilter}, SpeakerBeam \cite{zmolikova2019speakerbeam}, and bidirectional LSTM (BLSTM).

\subsubsection{Conformer}

The application of acquisition of time-frequency representation for Conformer-based TSE was proposed in~\cite{liu2024target}.  Conformer has a hybrid architecture, combining convolutional layers and multi-head attention mechanisms. This makes it effective in capturing both local and global features in the input audio.

Input Process: The input consists of a 256-dimensional real part and a 256-dimensional imaginary part of the STFT (obtained from a 512-point FFT with the DC component removed), along with a 192-dimensional x-vector (a speaker embedding extracted using ECAPA-TDNN and concatenated along the time dimension).

Extraction Network (Conformer blocks): The model consists of four stacked Conformer blocks:
\begin{itemize}
    \item Feedforward Layers: Each block begins with a feedforward layer of size 1024, with dropout set to 0.2.
    \item Multi-Head Attention: Multi-head attention is applied with four heads and a dropout rate of 0.2.
    \item Convolutional Layers: A convolution layer with a kernel size of 3, followed by batch normalization and the Swish activation function, is applied with dropout.
    \item Residual Connections: Residual connections with half-scaling are applied throughout the network.
    \item Final Linear Layer: The final output is a 512-dimensional feature vector, which is split into real and imaginary parts to compute the complex mask for STFT reconstruction.
\end{itemize}

Output Process: The model outputs a complex ratio mask~\cite{williamson2016complex} to apply to the complex-valued mixture STFT for reconstructing the target speaker's waveform.

\subsubsection{BLSTM}
 BLSTM  is a variation of the Conformer, in which the Conformer blocks have been replaced with BLSTM layers. Other aspects remain unchanged, except for the final fully connected layer, which is adjusted from 1024 to 512 dimensions to ensure consistent output.

\subsubsection{SpeakerBeam}
SpeakerBeam~\cite{zmolikova2019speakerbeam} is based on a BLSTM speaker encoder that extracts the target speaker’s voice using the speaker embedding as a guide. Different from the original SpeakerBeam , which uses magnitude spectra as input, we use complex spectra, as it has shown significant improvements over magnitude spectra. We use the same network architecture from the original version to extract speaker information. In such a case, only the input and output have been changed. This was done to evaluate whether our method is effective for the inner speaker information extraction network as well. Further details can be found in the complete architecture description in Appendix~\ref{arch_sb}.

\subsubsection{VoiceFilter}
Analogous to SpeakerBeam, only the input and output have been changed to complex spectra  for VoiceFilter, while the pre-trained d-vector used in the original version was replaced with the same x-vector as Conformer  . The rest of the structure remains the same.  Further details can be found in the complete architecture description in   Appendix~\ref{arch_vf}.

\subsection{Speaker information extraction model}
 ECAPA-TDNN~\cite{desplanques2020ecapa} is a state-of-the-art speaker encoder. It uses convolutional and residual blocks for feature extraction, followed by attentive statistical pooling \cite{astats_pooling} to generate speaker embeddings. The model is trained with additive angular margin softmax \cite{aam_softmax}. The original model, available via SpeechBrain \cite{speechbrain}, was trained on  VoxCeleb1~\cite{nagrani2017voxceleb} and VoxCeleb2~\cite{chung2018voxceleb2}.

For this study, we utilized the training framework from SpeechBrain, while retraining the model on the CN-Celeb dataset~\cite{li2022cn}. Although our target dataset contains English speech recordings, the model trained on CN-Celeb demonstrated superior performance compared to training on the original VoxCeleb 1+2 datasets. This improvement could be attributed to the increased speaker diversity and noise present in CN-Celeb.

\begin{figure*}[htbp]
    \centering
    \includegraphics[width=\textwidth]{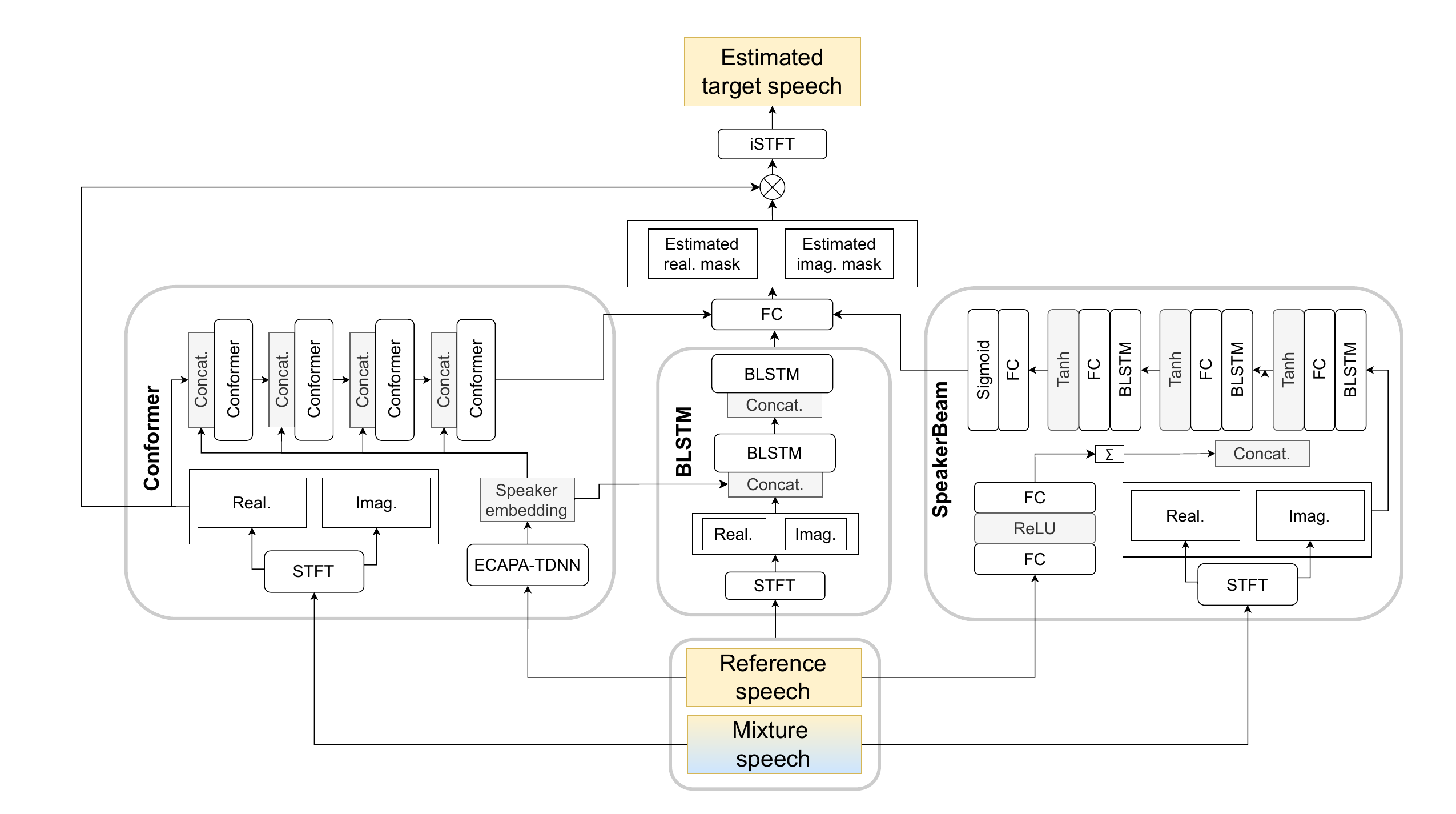}
    \caption{Details of Conformer, BLSTM, and SpeakerBeam TSE models.}
    \label{fig:alltsemodel}
\end{figure*}

\subsection{Loss function and evaluation metrics}
We use both  SNR as the loss function and  SDR as the evaluation metric to assess the quality of separated or enhanced speech in TSE tasks.

\subsubsection{Loss function}
The loss function used in this study is based on a negative SNR, calculated as the ratio between the power of the target speech and  error (difference between the target speech and   predicted target speech) in the time domain. It is expressed as:
\begin{equation}
\text{Loss}_{\textit{snr}} = -10 \log_{10}  \frac{ s^2}{ \left(s - \hat{s}\right)^2}, 
\end{equation}
A higher SNR indicates better reconstruction of the target speech. where $s$ represents the ground truth target speaker's speech, and $\hat{s}$ is the network's estimated target speaker's speech. The negative sign is used to convert the SNR into a loss value that can be minimized during training.

\subsubsection{Evaluation metric}
To evaluate the system  performance, we used the widely-used SDR metric~\cite{xu2020spex,xu2021target}, which measures the ratio of the target speech's power to the power of distortions introduced by the extraction system. For SDR computation, we use the implementation provided by \texttt{torchmetrics}~\cite{torchmetrics}.

\section{Experiments}

The goal with our experiments was to address two key research questions related to TSE using both real and synthetic data:

\begin{itemize}
    \item \textbf{Q1:} How effective is~\textit{Libri2Vox}   compared to    \textit{Libri2Talker}  in improving TSE performance?
    \item \textbf{Q2:} Does the use of synthetic speakers in    \textit{Libri2Vox-syn} dataset further enhance TSE performance, particularly in noisy environments, compared with using models trained only on real speakers?
\end{itemize}

To answer these questions, we conducted a series of experiments using the aforementioned TSE models trained on different combinations of real and synthetic data, with alternated noisy data augmentation. We evaluate these models on both the  \textit{Libri2Talker} and  \textit{Libri2Vox} test sets to comprehensively assess their performance across different conditions.
In addition to the main experiments, we conducted several ablation studies to explore the impact of specific components and configurations on the performance of the TSE models. These ablation studies were designed to provide further insights into the contributions of individual factors such as the inclusion of synthetic speakers,  ratio of real to synthetic data, and  effects of noise augmentation.

\subsection{Experimental setup and model configurations}

All TSE models were trained using a custom learning rate scheduler, designed to adjust the learning rate dynamically on the basis of the number of steps. Each step corresponds to one mini-batch, where the mini-batch size was set to 48. The initial learning rate was set to 1 × 10$^{-3}$, with a minimum threshold of 1 × 10$^{-5}$. The learning rate was warmed up linearly for the first 5000 steps (covering approximately 104,000 samples), after which it followed an inverse square root decay on the basis of the step count. Specifically, after the warm-up phase, the learning rate decayed proportionally to $(\text{warmup\_steps} / \text{global\_step})^{0.5}$. This dynamic learning rate schedule enabled for smooth transitions during training while avoiding rapid drops in learning rate that could destabilize the optimization process.

The Adam optimizer was used with the default settings of $\beta_1 = 0.9$, $\beta_2 = 0.999$, and $\epsilon = 10^{-8}$. No additional data augmentation, feature normalization, or input trimming was applied during training. All experiments were conducted on an Nvidia Tesla A100 GPU, with each model trained for three independent runs using different random seeds. The final results are reported as the average across these runs to ensure robustness and minimize the effects of random initialization.

The sampling frequency of the speech waveform was set to 16 kHz. The STFT parameters for the mixture signal included a window length of 32 ms and   hop size of 8 ms with a 512-point FFT.

\subsection{Training strategy}

\subsubsection{Noisy data augmentation}
We used noise from the DNS Challenge dataset~\cite{reddy2021icassp}. During training, there was a 50\% chance that a randomly selected 6-second noise segment would be dynamically mixed with the target speaker’s audio. The SNR for this dynamic mixing was uniformly sampled from the range of \([-5, 10]\) dB. The purpose of applying noise augmentation is to explore whether it can further enhance the model's performance, especially since   VoxCeleb2   already contains some real-world noise.

\begin{figure}[ht]
    \centering
    \includegraphics[width=0.45\textwidth]{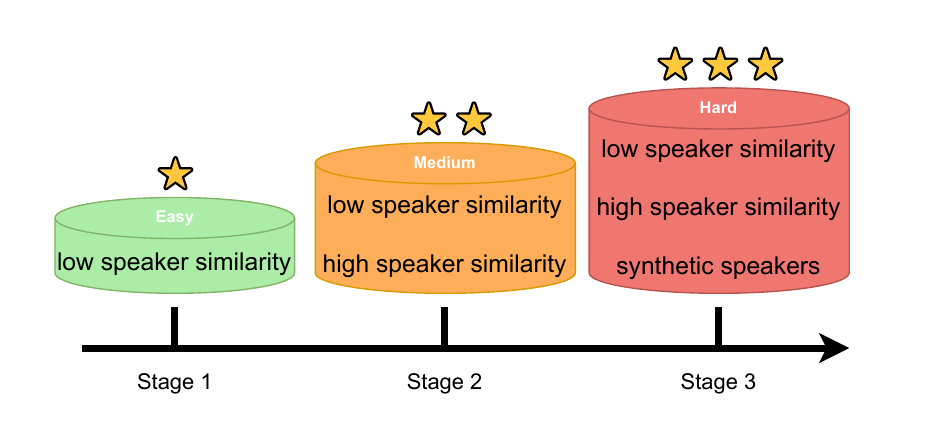}
    \vspace{-0.5cm}
    \caption{\centering Three stage curriculum learning.}
    \label{fig:3stage}
\end{figure}

\subsubsection{Curriculum learning}

To enhance the model’s ability to distinguish between target and interference speakers with varying degrees of similarity, we implemented  a CL strategy. Following our previous research\cite{liuslt}, the training process is divided into three stages, as illustrated in Fig. \ref{fig:3stage}:

\begin{itemize}
    \item \textbf{Stage 1:} In the first stage, the training data consists primarily of target and interference speaker pairs with low similarity. The goal at this stage is to enable the model to focus on simpler tasks, thus establishing a solid foundation for learning speaker characteristics.
    \item \textbf{Stage 2:} In the second stage, the model is exposed to speaker pairs with higher similarity, which gradually increases the complexity of the task.
    \item \textbf{Stage 3:} Finally, in the third stage, in addition to low and high similarity speaker pairs, synthetic interference speakers are introduced to further diversify the data and improve the model's generalization capability.
\end{itemize}

\begin{table*}[t]
\label{tab:libri2vox}
\centering
\caption{Experimental results for different models in iSDR(dB).}
\begin{tabular}{lcccrr}
\toprule
\textbf{Model} & \textbf{Libri2Talker} & \textbf{Libri2Vox} & \textbf{Noise aug.} & \textbf{Libri2Talker test set} & \textbf{Libri2Vox test set} \\
\midrule

% \multicolumn{6}{c}{\textbf{Conformer}} \\
% \midrule
\multirow{6}{*}{Conformer} 
 & $\checkmark$ & & & 9.13 & 8.45 \\
 & $\checkmark$ & & $\checkmark$ & 9.37 & 8.79 \\
 & & $\checkmark$ & & -1.28 & 15.42 \\
 & & $\checkmark$ & $\checkmark$ & -0.85 & 15.33 \\
 & $\checkmark$ & $\checkmark$ & & 9.69 & 14.31 \\
 & $\checkmark$ & $\checkmark$ & $\checkmark$ & 9.58 & 14.34 \\

\midrule

% \multicolumn{6}{c}{\textbf{BLSTM}} \\
% \midrule
\multirow{6}{*}{BLSTM} 
 & $\checkmark$ & & & 6.37 & 5.26 \\
 & $\checkmark$ & & $\checkmark$ & 6.19 & 6.06 \\
 & & $\checkmark$ & & -1.60 & 12.34 \\
 & & $\checkmark$ & $\checkmark$ & -1.34 & 13.18 \\
 & $\checkmark$ & $\checkmark$ & & 7.68 & 11.84 \\
 & $\checkmark$ & $\checkmark$ & $\checkmark$ & 7.52 & 12.25 \\
% \midrule

% \multicolumn{6}{c}{\textbf{SpeakerBeam}} \\
\midrule
\multirow{6}{*}{SpeakerBeam} 
 & $\checkmark$ & & & 6.07 & 5.40 \\
 & $\checkmark$ & & $\checkmark$ & 6.26 & 6.57 \\
 & & $\checkmark$ & & -1.57 & 11.62 \\
 & & $\checkmark$ & $\checkmark$ & -1.60 & 12.40 \\
 & $\checkmark$ & $\checkmark$ & & 7.32 & 11.47 \\
 & $\checkmark$ & $\checkmark$ & $\checkmark$ & 7.65 & 11.48 \\

\midrule
\multirow{6}{*}{Voicefilter} 
 & $\checkmark$ & & & 5.04  & 4.59 \\
 & $\checkmark$ & & $\checkmark$ & 4.57 & 4.85 \\
 & & $\checkmark$ & & -3.30 & 11.92 \\
 & & $\checkmark$ & $\checkmark$ & -2.86 & 11.63 \\
 & $\checkmark$ & $\checkmark$ & & 4.73 & 10.37  \\
 & $\checkmark$ & $\checkmark$ & $\checkmark$ & 4.32  &10.82   \\

\bottomrule
\end{tabular}
\label{t1}
\end{table*}

\section{Results and Discussion}

\subsection{Impact of Libri2Vox}
First, we discuss the conducted experiments to evaluate the effectiveness of the proposed Libri2Vox against the existing Libri2Talker. 
% Our analysis examined three state-of-the-art TSE models - Conformer, BLSTM, and SpeakerBeam - under various training configurations, including noise augmentation and cross-dataset combinations. 
The experimental results are presented in Table III. They revealed several significant findings regarding dataset-specific performance and cross-dataset benefits.
When training exclusively on Libri2Vox, we observed optimal performance on the Libri2Vox test set, achieving an iSDR of 15.42 dB with Conformer. However, this same configuration showed performance degradation (iSDR: -1.28 dB) when evaluated on the Libri2Talker test set, indicating domain mismatch between the datasets and thus the underlying complexity of the introduced dataset. Similarly, models trained solely on Libri2Talker exhibited degraded performance when tested on Libri2Vox, highlighting the importance of cross-dataset training strategies.

The joint training strategy using both Libri2Talker and Libri2Vox demonstrated remarkable improvements over single-dataset training, with Conformer achieving 14.31 dB iSDR on the Libri2Vox test set and 9.69 dB iSDR on the Libri2Talker test set. These results represent consistent improvements over single-dataset training across both test sets. 
Moreover, for such joint training case, compared to without it, the addition of noise augmentation shows consistent improvement on robustness across all exercised models, on Libri2Vox test set, with notably 0.41 dB improvement on BLSTM. 
% The addition of noise augmentation further enhanced model robustness, providing improvements of 0.24 dB for Libri2Talker-only training and 0.03 dB for combined dataset training.
These results indicate that attributed to the increased speaker diversity in Libri2Vox (over 7,000 speakers) and more realistic acoustic conditions inherited from real-world recordings from VoxCeleb2, Libri2Vox not only can serve as an effective standalone dataset but also can provide complementary benefits when combined with existing data.

\begin{table*}[t]
\centering
\caption{iSDR(dB) results of three stage methods. "Real only" means   use of only real data with cosine similarity less than 0.5 in  1st stage  (about \textit{71\%} of all data).}
\begin{tabular}{llccc|c}
    \toprule
    \textbf{Model} & \textbf{Method} & \textbf{Stage 1} & \textbf{Stage 2} & \textbf{Stage 3} & \textbf{Libri2Vox Test Set} \\
    \midrule
    \multirow{6}{*}{Conformer} & w/o CL (Real only) & \checkmark & & & 15.42 \\
    & w/o CL (Real+SynVox2) & \checkmark & & & 15.46 \\
    & w/o CL (Real+SALT) & \checkmark & & & 15.43 \\
    & w/ 1-stage CL (Real only) & \checkmark & & & 15.39 \\
    & w/ 2-stage CL (Real only) & \checkmark & \checkmark & & 15.89 \\
    & w/ 3-stage CL (Real only) & \checkmark & \checkmark & \checkmark & 16.01 \\
    & w/ 3-stage CL (Real + SynVox2) & \checkmark & \checkmark & \checkmark & \textbf{16.20} \\
    & w/ 3-stage CL (Real + SALT) & \checkmark & \checkmark & \checkmark & \textbf{16.20} \\
    \midrule
    \multirow{6}{*}{BLSTM} & w/o CL (Real only) & \checkmark & & & 12.34 \\
    & w/o CL (Real+SynVox2) & \checkmark & & & 12.53 \\
    & w/o CL (Real+SALT) & \checkmark & & & 12.54 \\
    & w/ 1-stage CL (Real only) & \checkmark & & & 11.73 \\
    & w/ 2-stage CL (Real only) & \checkmark & \checkmark & & 12.50 \\
    & w/ 3-stage CL (Real only) & \checkmark & \checkmark & \checkmark & 12.65 \\
    & w/ 3-stage CL (Real + SynVox2) & \checkmark & \checkmark & \checkmark & \textbf{13.07} \\
    & w/ 3-stage CL (Real + SALT) & \checkmark & \checkmark & \checkmark & 13.06 \\
    \midrule
    \multirow{6}{*}{SpeakerBeam} & w/o CL (Real only) & \checkmark & & & 11.62 \\
    & w/o CL (Real+SynVox2) & \checkmark & & & 12.01 \\
    & w/o CL (Real+SALT) & \checkmark & & & 12.07 \\
    & w/ 1-stage CL (Real only) & \checkmark & & & 11.17 \\
    & w/ 2-stage CL (Real only) & \checkmark & \checkmark & & 11.76 \\
    & w/ 3-stage CL (Real only) & \checkmark & \checkmark & \checkmark & 11.88 \\
    & w/ 3-stage CL (Real + SynVox2) & \checkmark & \checkmark & \checkmark & \textbf{12.35} \\
    & w/ 3-stage CL (Real + SALT) & \checkmark & \checkmark & \checkmark & 12.34 \\

    \midrule
    \multirow{6}{*}{Voicefilter} & w/o CL (Real only) & \checkmark & & & 11.92 \\
    & w/o CL (Real+SynVox2) & \checkmark & & & 11.41 \\
    & w/o CL (Real+SALT) & \checkmark & & & 11.46 \\
    & w/ 1-stage CL (Real only) & \checkmark & & & 11.34 \\
    & w/ 2-stage CL (Real only) & \checkmark & \checkmark & & 12.10 \\
    & w/ 3-stage CL (Real only) & \checkmark & \checkmark & \checkmark & 12.15 \\
    & w/ 3-stage CL (Real + SynVox2) & \checkmark & \checkmark & \checkmark & \textbf{12.39} \\
    & w/ 3-stage CL (Real + SALT) & \checkmark & \checkmark & \checkmark & 12.26 \\
    \bottomrule
\end{tabular}
\label{tab:main_syn_exp}
\vspace{-3mm}
\end{table*}

\subsection{Synthetic Libri2Vox}
We then conducted extensive experiments to evaluate the impact of synthetic data with CL on TSE performance, with results presented in Table~\ref{tab:main_syn_exp}. 
% Our experimental framework encompassed four models to systematically examined their behavior under various training configurations.
We implemented a progressive training strategy with three distinct stages. We evaluated eight configurations for each model, including baseline without CL, single-stage CL with real data, two-stage CL with real data, three-stage CL with real data, and two variants of three-stage CL incorporating synthetic data (SynVox2 or SALT). To isolate the impact of synthetic data from potential benefits of extended training, we included a controlled condition under which the third stage continued with real data only (w/ 3-stage CL (Real only)). The results clearly indicate that while additional training epochs with real data provided slight improvements, they were inferior to the gains achieved through synthetic data integration, validating the effectiveness of introducing the synthetic data into the training.

Results presented demonstrate consistent performance improvements across all models when implementing CL strategies. Conformer demonstrated particularly notable gains, with iSDR improving from a baseline of 15.42 to 15.89 dB using three-stage CL with real data. The introduction of synthetic data in the third stage further enhanced performance, achieving 16.20 dB with both SynVox2 and SALT variants. Importantly, this improvement surpassed the control condition using only real data (16.01 dB), confirming that the benefits stem from the synthetic data rather than extended training epochs. 

To demonstrate the benefits of CL in utilizing synthetic data, starting with 50\% real and 50\% synthetic data at stage 1 only (e.g., w/o CL (Real+SynVox2)) does not show substantial improvements compared to using synthetic and real data in Stage 3 (e.g., w/ 3-stage CL (Real + SynVox2)), while curriculum learning further enhances the performance of the latter setup.

% \textcolor{red}{Interestingly, we also see that performance degrades when training using only the synthetic data and without CL. CL is effective when using the synthetic data.}(except for the VoiceFilter model)

While BLSTM, SpeakerBeam, and Voicefilter exhibited lower absolute performance compared with  Conformer, they showed proportionally larger relative gains from CL implementation. BLSTM achieved a 0.73 dB improvement with synthetic data integration, while SpeakerBeam and Voicefilter showed gains of 0.73 and 0.47 dB respectively. Notably, both synthetic data  methods (SynVox2 and SALT) yielded comparable improvements, likely due to their shared use of the HiFi-GAN vocoder architecture. These results collectively indicate the effectiveness of combining  synthetic data augmentation with CL  for enhancing TSE system performance.

\subsection{Ablation study}

\begin{figure}[ht]
    \centering
    \includegraphics[width=0.45\textwidth]{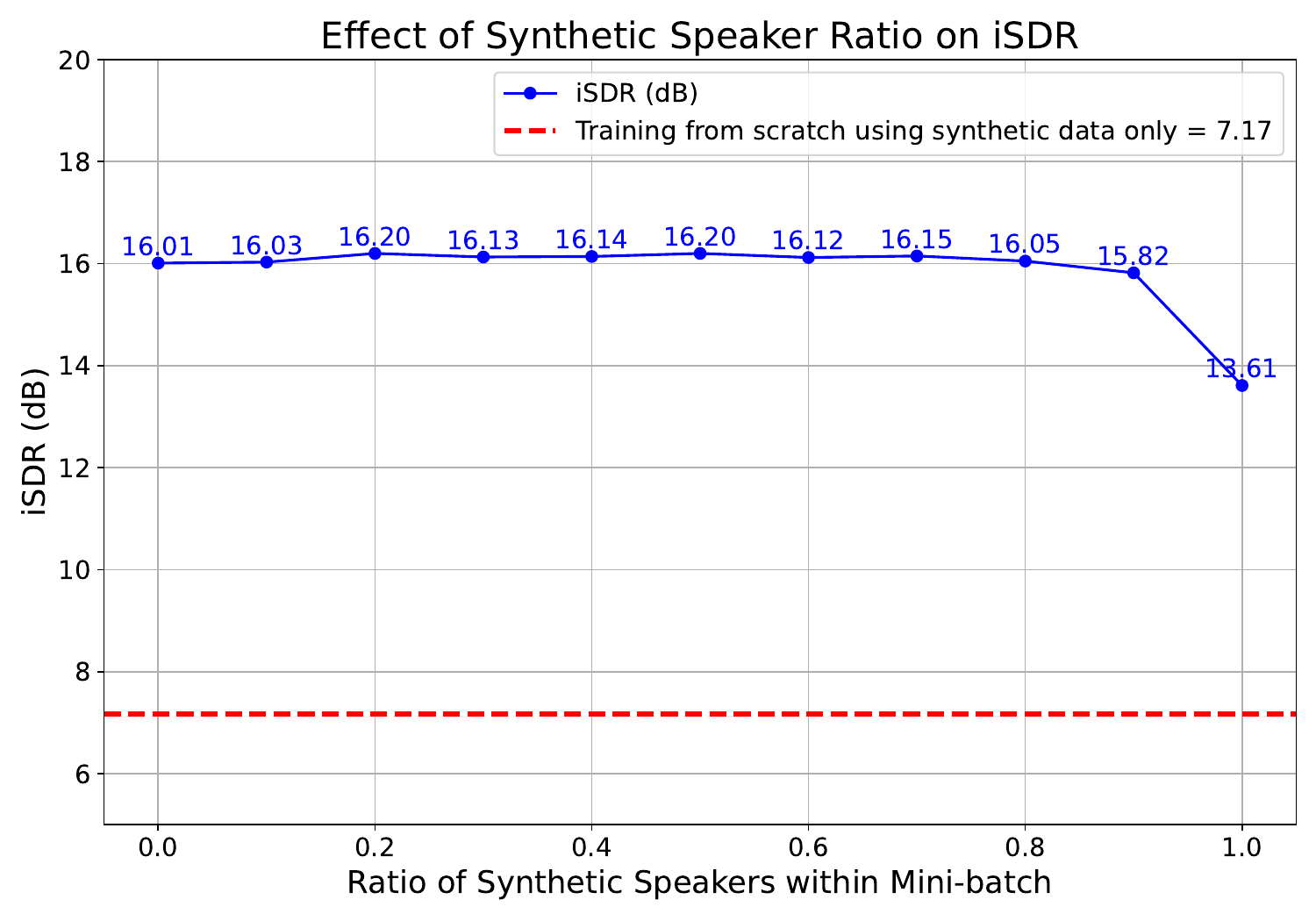}
    \caption{Impact of synthetic speaker ratio within one batch at Stage 3 of the configuration ``w/ 3-stage CL (Real + SALT)" on Conformer. The red dashed line corresponds to the performance (7.17 dB) where the training starts from scratch using synthetic data only.}
    \label{fig:synratio}
\end{figure}

\subsubsection{Comparison of different synthetic speakers ratios within a mini-batch}
As shown in Fig. \ref{fig:synratio}, we evaluated the impact of different ratios of synthetic speakers within a mini-batch on the performance of Conformer. The experiments involved the Stage 3 architecture with SALT-generated synthetic speakers, following the configuration ``w/ 3-stage CL (Real + SALT)" shown in Table~\ref{tab:main_syn_exp}.
Optimal performance was achieved with synthetic speaker ratios of 0.2 and 0.5, both yielding an iSDR of 16.20 dB. This configuration significantly outperformed the baseline configuration using only real data (ratio = 0.0), which achieved an iSDR of 16.01 dB. However, increasing the synthetic speaker ratio beyond these optimal values led to performance degradation. Note that when synthetic speakers comprised 90\% of the training data, the iSDR decreased to 15.82 dB. Such degradation was furthered with the ratio being 1.0, resulting in an iSDR of 13.61 dB.

It is worth mentioning that as shown via the red dashed line in the figure, if we start training the model from scratch using only synthetic data instead of using the aforementioned configuration, the resulting iSDR is 7.17 dB. This poor result is due to the significant mismatch between the synthetic training data and  real test data. Therefore, implementing the CL method with which the model is first trained on real data before progressively incorporating synthetic data, proves to be substantially more effective.

\subsubsection{Comparison of different numbers of synthetic speakers}
\begin{table}[t]
%\label{tab:num_spk}
\caption{iSDR(dB) results of different numbers of synthetic speakers for BLSTM. ``\textbf{No.}" is the experimental index for convenience of description.}
\centering
\begin{tabular}{c|c|c|c}
\hline
\textbf{No.} & \textbf{Stage3}      & \textbf{Stage4}        & \textbf{iSDR} \\ \hline
1            & SynVox2              & ---                    & 13.07         \\ \hline
2            & SALT                 & ---                    & 13.06         \\ \hline
3            & SynVox2 + SALT       & ---                    & 13.16         \\ \hline
4            & SALT                 & SynVox2                & \textbf{13.44}         \\ \hline
5            & SynVox2              & SALT                   & 13.43         \\ \hline
6            & SynVox2              & SynVox2 + SALT         & 13.30         \\ \hline
%7            & synvox2              & real                   & 13.17         \\ \hline
%8            & real                 & ---                    & 12.65         \\ \hline
%9            & real                 & real                   & 12.73         \\ \hline
\end{tabular}
\label{tab:num_spk}
\end{table}

The goal of this additional study is to determine if increasing the number of synthetic speakers would improve the performance of BLSTM for the 3-stage CL. Table~\ref{tab:num_spk} presents the experimental results across different synthetic speaker configurations. Given that SynVox2 and SALT generate distinct speaker sets, their combination effectively increases the total number of synthetic speakers. Our experiments focused on Stage 3 and Stage 4 implementations, while maintaining consistent configurations for Stages 1 and 2 as specified in Table~\ref{tab:main_syn_exp} under "w/ 2-stage CL (Real only)". Specifically, Stage 1 utilized data with cosine similarity below 0.5, and Stage 2 incorporated the complete dataset. For example, Experiment 4 involved using SALT for Stage 3 and additionally SynVox2 for Stage 4.

When both synthetic datasets were used simultaneously in Stage 3 (\textbf{No.} 3), we observed an improvement to 13.16 dB, indicating the potential benefits of increased speaker diversity.
However, the most significant performance gains emerged from sequential training strategies. \textbf{No.} 4 and 5 implemented alternating synthetic datasets between Stage 3 and Stage 4, achieving the highest iSDR (13.44 dB and 13.43 dB respectively). Such sequential approach proved more effective than simultaneous dataset utilization (\textbf{No.} 6, 13.30 dB), suggesting that progressively introducing different synthetic speaker sets substantially enhance TSE performance.

\section{Conclusion}

In this paper, we introduced Libri2Vox, a novel dataset designed to address the challenges of TSE in real-world acoustic environments. The dataset combines clean speech from LibriTTS with naturally noisy interference from VoxCeleb2, creating a diverse training environment with over 7,000 speakers. We further enhanced the dataset's utility through synthetic data generation, developing two complementary methods, SynVox2 and SALT, to expand speaker diversity. Our comprehensive evaluation demonstrated that combining synthetic speakers with CL significantly improved model performance.

While synthetic speaker data offers valuable diversity for training TSE models, it is important to recognize the inherent limitations of using speech generative models. The useful distribution provided with these generative models is often constrained, and understanding which specific distributions are most beneficial for training remains a challenge. In some cases, only a small portion of the generated data may be truly useful for covering the necessary distribution. Generating more data beyond this point might not provide additional value, as certain types of data become redundant, offering no new information for the model~\cite{gan2024towards}. The ``new knowledge" provided with these speech generative models can quickly become repetitive, and the model may not need to repeatedly learn from similar data.
% ,wang2024do

For future work, we will explore what types of synthetic data are most beneficial for TSE, potentially through data selection methods. For example, tracking the gradient changes of a network at each step could help determine whether the current data is useful for training. 
We will also explore using dataset distillation methods to generate synthetic data that represents real data characteristics. These synthetic data points could help us better understand the types of representations needed for TSE tasks.

\bibliographystyle{IEEEtran}
\bibliography{main}

\balance

\appendix

\begin{appendices}
\section{Details of network architecture}

\subsection{Architecture of SpeakerBeam}
\label{arch_sb}
\textbf{Input Process}: The real and imaginary parts of the STFT are concatenated, resulting in a 512-dimensional input.  The speaker embedding is extracted by processing the magnitude of the STFT of the reference speech through a fully connected layer  then concatenated along the time dimension, resulting in an input of $T \times (256 + 256 + 192)$.

\textbf{BLSTM Layers}: This model uses 3 BLSTM layers, each with 512 hidden units per direction. The first layer processes the concatenated input.

\textbf{Fully Connected Layers}: Each BLSTM layer's output is passed through fully connected layers, reducing the output back to 512 dimensions.

\subsection{Architecture of Voicefilter}
\label{arch_vf}
\textbf{Convolutional Layers}: This model comprises 8 convolutional layers:
\begin{itemize}
    \item Layers are zero-padded with kernel sizes varying from 1 to 7 and 64 output channels.
    \item Dilation factors increase from 1 to 16, and each layer is followed by batch normalization and ReLU activation.
    \item The final layer is a 2-channel convolution with a kernel size of 1.
\end{itemize}

\textbf{LSTM Layer}: A BLSTM with 400 hidden units per direction processes the concatenated convolution outputs and the 192-dimensional x-vector, which, similar to  Conformer, is concatenated along the time dimension before being fed into the LSTM.

\textbf{Fully Connected Layers}: The first fully connected layer reduces the LSTM output to 600 dimensions. The second fully connected layer maps the result to 512 dimensions (for real and imaginary parts of the STFT).

\end{appendices}

% \renewcommand{\thetable}{A\arabic{table}}
% % \include{appendix.tex}

% % \vfill

\end{document}